\documentclass[12pt,twoside]{article}
\usepackage[mathscr]{eucal}
\usepackage{amsmath,amsfonts,amssymb,amsthm}
\usepackage[matrix,arrow]{xy}
\usepackage{mathabx}
\usepackage{amsthm,amsmath,latexsym,amssymb,amsfonts,amsbsy}
\usepackage{wrapfig}
\usepackage{graphicx}
\usepackage{float}
\usepackage{soul}

\usepackage[breaklinks]{hyperref}
\usepackage{amsmath}
\usepackage{amssymb}
\usepackage{braket}
\usepackage[T1]{fontenc}
\usepackage{sidecap}

\usepackage{tikz,pgf}
\usetikzlibrary{shapes}
\usetikzlibrary{calc}
\usetikzlibrary{decorations.pathmorphing}
\usetikzlibrary{decorations.pathreplacing,shapes.misc}
\usetikzlibrary{positioning}
\usetikzlibrary{arrows}
\usetikzlibrary{decorations.markings}
\usetikzlibrary{shadings}

\usetikzlibrary{intersections}

\def\be{\begin{equation}}
\def\ee{\end{equation}}
\def\ba{\begin{array}}
\def\ea{\end{array}}

\def\dps{\displaystyle}
\newcommand{\half}{\frac{1}{2}}

\def\1{\tilde{1}}
\def\2{\tilde{2}}
\def\3{\tilde{3}}

\newdimen\tableauside\tableauside=1.0ex
\newdimen\tableaurule\tableaurule=0.4pt
\newdimen\tableaustep
\def\phantomhrule#1{\hbox{\vbox to0pt{\hrule height\tableaurule
width#1\vss}}}
\def\phantomvrule#1{\vbox{\hbox to0pt{\vrule width\tableaurule
height#1\hss}}}
\def\sqr{\vbox{%
  \phantomhrule\tableaustep

\hbox{\phantomvrule\tableaustep\kern\tableaustep\phantomvrule\tableaustep}%
  \hbox{\vbox{\phantomhrule\tableauside}\kern-\tableaurule}}}
\def\squares#1{\hbox{\count0=#1\noindent\loop\sqr
  \advance\count0 by-1 \ifnum\count0>0\repeat}}
\def\tableau#1{\vcenter{\offinterlineskip
  \tableaustep=\tableauside\advance\tableaustep by-\tableaurule
  \kern\normallineskip\hbox
    {\kern\normallineskip\vbox
      {\gettableau#1 0 }%
     \kern\normallineskip\kern\tableaurule}%
  \kern\normallineskip\kern\tableaurule}}
\def\gettableau#1 {\ifnum#1=0\let\next=\null\else
  \squares{#1}\let\next=\gettableau\fi\next}

\renewcommand{\tilde}{\widetilde}
\renewcommand{\hat}{\widehat}

\newcommand{\bref}[1]{\textbf{\ref{#1}}}

\newcommand{\im}{\mathop{\mathrm{Im}}}

\def\cO{\mathcal{O}}

\def\cV{\mathcal{V}}

\numberwithin{equation}{section} \makeatletter
\@addtoreset{equation}{section}

\def\be{\begin{equation}}
\def\ee{\end{equation}}
\def\ba{\begin{array}}
\def\ea{\end{array}}

\def\dps{\displaystyle}

\def\ba{\begin{array}}
\def\ea{\end{array}}

\def\dps{\displaystyle}

\def\tepsilon{\tilde \epsilon}
\def\tDelta{\tilde\Delta}

\begin{document}

\begin{flushright}
FIAN-TD-2016-06 \\
\end{flushright}

\vspace{1cm}

\begin{center}

{\Large\textbf{Holographic  interpretation of 1-point toroidal block
\\
[4mm]
in the semiclassical limit}}

\vspace{5mm}

\vspace{.9cm}

{\large K.B. Alkalaev$^{\;a,b}$ and   V.A. Belavin$^{\;a,c}$}

\vspace{0.5cm}

\textit{$^{a}$I.E. Tamm Department of Theoretical Physics, \\P.N. Lebedev Physical
Institute,\\ Leninsky ave. 53, 119991 Moscow, Russia}

\vspace{0.5cm}

\textit{$^{b}$Moscow Institute of Physics and Technology, \\
Dolgoprudnyi, 141700 Moscow region, Russia}

\vspace{0.5cm}

\textit{$^{c}$Department of Quantum Physics, \\ 
Institute for Information Transmission Problems, \\
 Bolshoy Karetny per. 19, 127994 Moscow, Russia}

\vspace{0.5cm}

\thispagestyle{empty}


\end{center}
\begin{abstract}

We propose the holographic interpretation of the 1-point conformal block on a torus in the semiclassical regime. 
To this end we consider the linearized  version of the block and  find its coefficients by means  of the perturbation procedure around natural seed configuration corresponding to the zero-point block. From the AdS/CFT perspective the linearized block is given by the geodesic length of the tadpole graph embedded into the thermal AdS plus the holomorphic part of the thermal AdS action. 

\end{abstract}

\section{Introduction}

The AdS$_3$/CFT$_2$ correspondence is   rich and diverse subject which is in a constant state of growth. 
In particular, many exact results  are known  in the semiclassical regime where the central charge tends to infinity or, equivalently, the gravitational coupling is  small \cite{Brown:1986nw}. In particular, it has been   shown recently that  classical conformal blocks on the Riemann sphere  are identified with lengths of geodesic networks stretched in the asymptotically  AdS space with  the angle deficit or BTZ black hole \cite{Hartman:2013mia,Fitzpatrick:2014vua,Asplund:2014coa,Caputa:2014eta,Hijano:2015rla,Fitzpatrick:2015zha,Perlmutter:2015iya,Alkalaev:2015wia,Hijano:2015qja,Banerjee:2016qca}.

Meanwhile, it is well known that the general  solution to the classical Euclidean  AdS$_3$ gravity is topologically associated with a solid torus \cite{Carlip:1994gc},  so that the corresponding boundary CFT is essentially toroidal theory characterized by parameters of the given particular solution 
\cite{Maloney:2007ud}. 
In this paper we are interested in toroidal conformal blocks and their dual realization.
For the previous studies of the 
toroidal conformal blocks in the framework of  CFT  see \cite{Fateev:2009me,Poghossian:2009mk,Hadasz:2009db,Fateev:2009aw,Menotti:2010en,Marshakov:2010fx,KashaniPoor:2012wb,Piatek:2013ifa}.

We propose the following holographic interpretation of the linearized classical 1-point  block on a torus. 
 The bulk geometry is identified  with the thermal AdS, while both intermediate and external fields of the classical block are represented by propagating massive particles with masses given by  classical conformal dimensions. Note that in the toroidal case the  background is not produced by  fields of the 1-point function, both the external and intermediate particles  are dynamical. This is in contrast with conformal blocks on the Riemann sphere appeared in the AdS/CFT context, where two heavy fields create  singularities of the corresponding angle deficit/BTZ geometries.
It is clear that the presence of the heavy fields in that case was aimed to produce a cylindrical topology for the boundary CFT which is appropriate for the consideration in the AdS/CFT
context.  

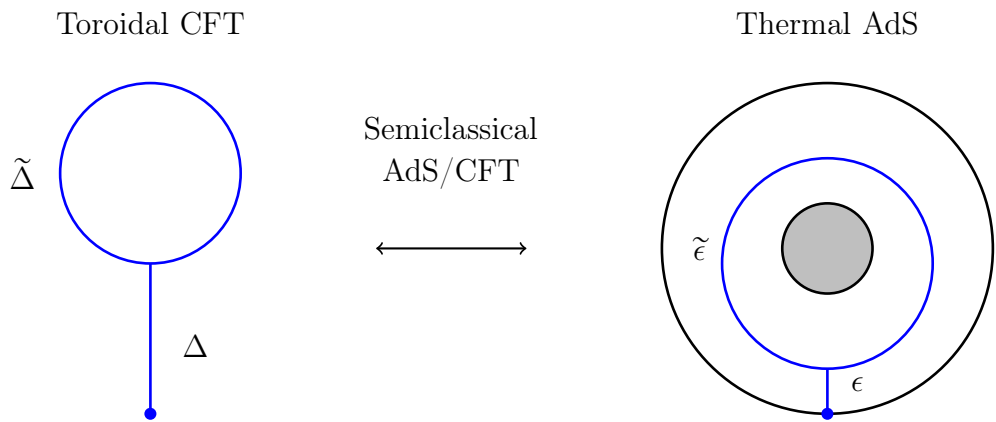
\begin{figure}[H]
\centering

\begin{tikzpicture}[line width=1pt]


\draw[blue] (-9,0) circle (1.2cm);
\draw[blue] (-9,-1.2) -- (-9,-3.2);
\fill[blue] (-9,-3.2) circle (0.8mm);

\draw (-10.7,0) node {$\tDelta$};
\draw (-8.4,-2.3) node {$\Delta$};

\draw (-9,2) node {$\text{Toroidal CFT}$};



\draw (-5,0.6) node {$\text{Semiclassical}$};
\draw (-5,0) node {$\text{AdS/CFT}$};


\draw[thick,<->] (-6,-1.0) -- (-4,-1.0);



\draw (0,2) node {$\text{Thermal AdS}$};


\draw[fill=lightgray] (0,-1.0) circle (0.6cm);
\draw (0,-1.0) circle (2.2cm);

\draw[blue] (0,-1.2) circle (1.4cm);
\draw[blue] (0,-2.6) -- (0,-3.2);
\fill[blue] (0,-3.2) circle (0.8mm);

\draw (-1.7,-1.0) node {$\tepsilon$};
\draw (0.4,-2.8) node {$\epsilon$};

\end{tikzpicture}
\caption{One-point conformal block realized as the tadpole graph embedded into the thermal AdS. The loop of the conformal block graph is identified with the non-contractible circle of the thermal AdS. $\Delta$ and $\tDelta$ are external and intermediate conformal dimensions,  $\epsilon = k \Delta$ and $\tepsilon = k \tDelta$ are classical conformal dimensions ($k=c/6$).} 
\label{duality}
\end{figure}

The main result can be formulated as follows. We find that modulo regulator dependent (infinite) terms  the linearized version of  1-point classical block function $f^{lin}$ with  $\epsilon, \tepsilon$ being external and intermediate classical conformal dimensions is  given by  
\be
\label{identification}
-f^{lin} = S_{thermal}+ \tepsilon\, S_{loop}  + \epsilon\, S_{leg}\;,
\ee 
where the first term is the holomorphic part of the  3d gravity action evaluated on the thermal AdS space, while the second and third terms give the length of the tadpole graph attached at some boundary point, see  Fig. \bref{duality}.

The outline of this paper is as follows. In section \bref{sec:1ptblock} we introduce the 1-point  block on a torus, discuss its classical limit and
then define linearized version of  the block. In general, the definition requires introducing  certain hierarchy of the conformal dimensions. In the 
case under consideration it can be described by the  ratio of the external and internal conformal dimensions of the fields.  We  describe the series expansion
of the block in terms of this parameter. In section \bref{sec:dual}  we  develop the holographic interpretation. In particular, in section \bref{sec:perturbative} we implement the perturbative procedure leading to the expansion of the geodesic length  analogous to those obtained  in section \bref{sec:1ptblock} for the conformal block. We discuss the seed geodesic configuration and its deformation, and find out that the length of the tadpole graph reproduces the conformal block function. We conclude in section \bref{sec:conclusion} by discussing future perspectives. Appendix \bref{sec:appA} contains the block coefficients represented as formal power series in the elliptic parameter.  

\section{Classical 1-point toroidal block}
\label{sec:1ptblock}

By analogy with CFT on the plane, 1-point correlation functions on the torus can be decomposed into conformal blocks (see, {\it e.g.}, \cite{Hadasz:2009db} for a review). The (holomorphic) conformal block of the 1-point correlator of the primary field $\phi_{\Delta}$  is given by 
\be
\label{Virtor}
\cV(\Delta, \tilde\Delta, c|q) = q^{\tilde \Delta  - c/24}\sum_{n = 0}^{\infty} q^n \cV_{n}(\Delta, \tilde\Delta, c)\;,
\ee  
where 
\be
\label{qtau}
q = e^{2\pi i \hat\tau}\;,  
\ee
is the elliptic parameter on a torus with the modulus $\hat\tau$, $\tDelta$ is the conformal dimension of the intermediate channel, and the expansion coefficients are
\be
\label{1ptcoef}
\cV_{n}(\Delta, \tilde\Delta, c) = \frac{1}{\langle \tilde \Delta| \phi_{\Delta} | \tilde \Delta\rangle }\sum_{n = |M|=|N|} G^{-1}_{MN}\,\langle \tilde \Delta, M| \phi_{\Delta} |N, \tilde \Delta\rangle\;.
\ee
Here, $| \tilde \Delta, M\rangle$ are the $M$-th level descendant vectors in the Verma module generated from the primary state $|\tilde \Delta\rangle$,  $G^{-1}_{MN}$ are elements of the inverse Gram matrix,  $G_{MN} = {\langle \tilde \Delta, M |N, \tilde \Delta\rangle}$. As usual, $|M|$ denotes the minus sum of Virasoro generator indices. Note that the 1-point conformal block is independent of the insertion point $z$.

One-point classical  block on a torus  can be defined along the same lines as in the spherical case  as the leading contribution in the $c\to \infty$ approximation \cite{Piatek:2013ifa}
\be
\label{ccb}
\cV(\Delta, \tilde \Delta|q) \sim   \exp\big[-k f(\epsilon, \tilde \epsilon|q)\big]\;,
\ee
where  $k=c/6$, external and intermediate classical conformal dimensions $\epsilon$, $\tilde \epsilon$  are related to the conformal dimensions  by 
\be
\label{deltas}
\Delta = k \epsilon \;,\qquad\;\; \tilde \Delta = k \tepsilon\;,
\ee
and the  function $f(\epsilon, \tilde \epsilon|q)$ is the proper classical one-point conformal block. The series expansion of this  function  can be easily computed by virtue of the  Virasoro algebra commutation relations 
\be
\label{class_block}
f(\epsilon, \tilde \epsilon|q) = (\tilde\epsilon - 1/4) \log q + \sum_{n=1}^\infty q^n \text{f}_n (\epsilon, \tilde \epsilon)\;,
\ee
where the first few coefficients are given by 
\be
\ba{c}
\label{coef}
\dps
\text{f}_1 = \frac{\epsilon^2}{2\tilde\epsilon}\;,
\qquad \;\;
\text{f}_2=\frac{\epsilon^2[24 \tilde\epsilon^2(4\tilde \epsilon+1)+\epsilon^2(5\tilde\epsilon-3) - 48 \epsilon \tilde \epsilon^2]}{16 \tilde \epsilon^3(4\tilde \epsilon +3)}\;, 
\\
\\
\dps
\text{f}_3 =\frac{\epsilon^2[\epsilon^4 (9 \tilde\epsilon^2-19 \tilde\epsilon+6)-16 \epsilon^3 \tilde\epsilon^2 (7 \tilde\epsilon-6)+8 \epsilon^2 \tilde\epsilon^2 (60 \tilde\epsilon^2+7 \tilde\epsilon-6)-192 \epsilon \tilde\epsilon^4 (4 \tilde\epsilon+3)+48 \tilde\epsilon^4 (8 \tilde\epsilon^2+10 \tilde\epsilon+3)]}{48 \tilde\epsilon^5 (4 \tilde\epsilon^2+11 \tilde\epsilon+6)}\;.\ea
\ee
(Coefficients $\text{f}_1$ and $\text{f}_2$ can also be found in \cite{Piatek:2013ifa}.)


The 1-point linearised classical block on a torus can be  defined as in the spherical case \cite{Hijano:2015rla,Alkalaev:2015wia,Alkalaev:2015lca,Alkalaev:2015fbw}.
To this end we introduce the lightness parameter 
\be
\label{lightness}
\delta = \frac{\epsilon}{\tilde\epsilon}<1  \;.
\ee
Then, changing from $(\epsilon, \tepsilon)$ to $(\delta,\tepsilon)$ we represent the  classical conformal block \eqref{class_block}  as a double series expansion in $q$ and $\delta$ keeping terms at most linear in $\tepsilon$,  
\be
f(\epsilon,\tilde{\epsilon}|q) = f^{lin}(\delta,\tilde\epsilon|q)+\mathcal{O}(\tilde\epsilon^2)\;,
\ee
where $f^{lin}(\delta,\tilde\epsilon|q)$ is the linearized classical block. Recalling  the definition \eqref{class_block} we see that the linearized block is given by    
\be
\label{vovalin}
f^{lin}(\delta,\tilde\epsilon|q)=(\tilde\epsilon - 1/4) \log q + \tepsilon \sum_{n=1}^{\infty} f_n(q) \delta^{2n}\;, 
\ee
where the  first  few coefficients are given  in  \eqref{appcoef1}-\eqref{appcoef2} as power series in $q$. The expansion coefficients can be written  in a closed form  as  
\begin{align}
&f_1(q)= \frac{q}{2}\frac{1}{1-q}\;,\qquad f_2(q)= \frac{q^2}{48}\frac{(q-3)}{(1-q)^3}\;,
\qquad f_3(q)=\frac{q^3}{480}\frac{q^2-5 q+10}{(1-q)^5}\;,\\
&f_4(q)=\frac{q^4}{3584} \frac{q^3-7 q^2+21 q-35}{(1-q)^7}\;,\quad 
f_5(q)=\frac{q^5}{23040} \frac{q^4-9 q^3+36 q^2-84 q+126}{(1-q)^9}\;,\quad \dots \;,
\end{align}
where the general element is given by 
\be
f_n = \frac{q^n}{\varkappa_n} \,\frac{q^{n-1}+ \gamma_{n-2} q^{n-2}+...+ \gamma_{1}}{(1-q)^{2n-1}}\;,
\ee
where $\gamma_i = (-)^i\binom{2n+i-1}{i}$, $i = 0,1,...,n-1$ are binomial coefficients, and $\varkappa_i$ are some constants. 

Let us note that  setting $\epsilon =0$ we arrive at the 0-point block which is the   Virasoro character.  It follows that the 1-point  block can be  considered as a small deformation of the 0-point block where the deformation parameter is identified with the external dimension. Such a deformation procedure is exactly what we have for the higher-point  conformal blocks on a sphere, where $n$-point classical block was  treated as a deformation of $(n-1)$-point block \cite{Alkalaev:2015lca,Alkalaev:2015fbw}. The procedure effectively works when $(n-1)$-point block is exactly known while the $n$-point block is not. This is the case with 1-point toroidal block, where 0-point block is the known character.

\section{Dual interpretation}
\label{sec:dual}

\begin{figure}[H]
\centering
\begin{tikzpicture}[line width=1pt]
\draw (0,0) circle (3.5cm);
\draw[fill=lightgray] (0,0) circle (1cm);

\draw[dashed] (0,0) circle (2.2cm);
\draw[blue] (0,-0.2) circle (2.43cm);

\draw[blue] (0,-2.65) -- (0,-3.5);


\fill[blue] (0,-3.5) circle (0.8mm);
\fill[blue] (0,-2.65) circle (0.8mm);

\draw (-2.0,2) node {$\tepsilon$};
\draw (-0.4,-3.0) node {$\epsilon$};

\draw (-0,-4.0) node {$\pi$};
\draw (-0,4.0) node {$0$};
\draw (-4.2,0) node {$\dps\frac{3\pi}{2}$};
\draw (4.2,0) node {$\dps\frac{\pi}{2}$};

\end{tikzpicture}
\caption{Annulus and tadpole graph. The inner and outer black solid circles represent the conformal boundary. The dashed circle goes along the $r=0$ radius. The blue loop is a deformation of the dashed circle when the external field represented by the solid blue segment is switched on.  Vertex  and boundary attachment points are at  $t=\pi$. Routinely,  time flows clockwise.} 
\label{annulus}
\end{figure}
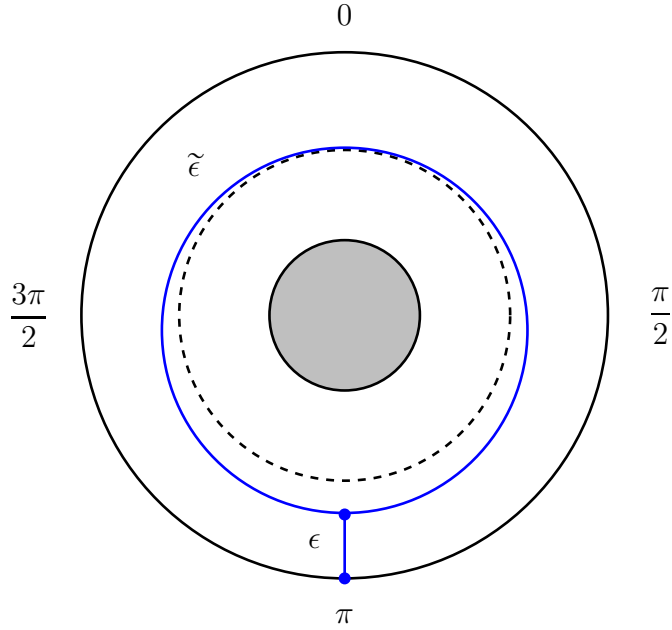

The linearized toroidal conformal block has the holographically dual interpretation where the block function is identified with the  length of  the  tadpole geodesic graph embedded into the three-dimensional bulk space, see Fig. \bref{duality} and \bref{annulus}. The tadpole is drawn on a two-dimensional annulus which is a slice of Euclidean  thermal AdS space  
\be
\label{thermal}
ds^2 = - \tau^2\, \big(1+ \frac{r^2}{l^2}\big)dt^2 +\big(1+ \frac{r^2}{l^2}\big)^{-1}dr^2 + r^2 d \varphi^2\;,
\ee
where $\tau$ is the pure imaginary modular parameter, and coordinates  $t \sim t + 2\pi$, $\varphi \sim \varphi + 2\pi$, $r \geq 0$. Topologically, the thermal AdS is a solid torus with  time running along the  non-contractible cycle. In what follows, we set the AdS radius $l=1$. 

Within the geodesic approximation the gravity functional integral is to be evaluated near the saddle-point given by a particular solution. It is known that in the low-temperature regime corresponding to $\im \tau \gg 1$ the  thermal AdS dominates the functional integral \cite{Maldacena:1998bw}. It follows that the on-shell classical action  for the gravity plus the matter represented by massive external and intermediate particles  is given by 
\be
\label{total}
S_{total} = S_{thermal} + \tepsilon\, S_{loop}+ \epsilon\, S_{leg}\;,
\ee
where the first term being the gravity  action  on the thermal AdS,  the second and third terms being the geodesic lengths of the loop and the radial line with conformal weights \eqref{deltas} identified in the $c\to \infty$ limit with masses. Below we show that that the linearized conformal block \eqref{vovalin} is equal to the total action \eqref{total} as $f^{lin} = -S_{total}$, cf. \eqref{identification}.  

In the sequel we calculate  the geodesic lengths within the classical mechanics of the external and intermediate particles propagating on the surface level characterized by a constant angle. However,  the spherical coordinates in \eqref{thermal} are incomplete in the sense that a loop going along the zeroth radius has no particular angle value. To complete the definition we change the parameterization of radial and angle coordinates as $-\infty< r < \infty$ and $\varphi \sim \varphi+ \pi$. Then, the solid torus is represented as a stack of annuluses on Fig. \bref{annulus} rotated along the $r=0$ circle by angles from $0$ to $\pi$. The $r=0$ circle has now definite though infinitely  degenerate angle position coinciding with that of the corresponding annulus. 


Let us first consider the case when the external field is the identity having $\Delta =0 $. Then, the 1-point conformal block becomes the 0-point conformal block identified with the Virasoro character.  The corresponding graph is a loop represented by a constant radius circle going along the origin $r = 0$, see Fig. \bref{annulus}. The total action \eqref{total} is reduced to 
$S_{total} = S_{thermal} + \tepsilon\, S_{loop}$. Recalling that the holomorphic part of  the gravitational action on the thermal AdS has been calculated in \cite{Carlip:1994gc,Maldacena:1998bw,Kraus:2006wn} to be $S_{thermal} = i\pi \tau/2$ in terms of the rescaled central charge $k$, while the circumference  is $S_{loop} = -2\pi i \tau$ we find that 
\be
\label{classchar}
S_{total}\Big|_{\epsilon=0} =  - 2 \pi i \, \left(\tilde \epsilon - 1/4\right)\,\tau \;:= \; -(\tilde \epsilon - 1/4) \log q =- f^{lin}(\epsilon, \tepsilon|q)\Big|_{\epsilon=0} \;.
\ee 
This  is exactly the minus leading term in the classical block \eqref{vovalin} provided the modular parameters are equated.


If the external field is switched on  $\epsilon \neq 0$ then the constant radius circle identified with the 0-point block is  deformed by the external field leg stretching  from the conformal boundary to some vertex point in the bulk. \footnote{In the region $\delta>1$ the loop is to be considered as a perturbation to the external leg so that in the limit $\delta = \infty$ the loop decouples from the leg because $\tepsilon = \epsilon/\delta$. On the CFT side, the classical block coefficients diverge in this limit, cf. \eqref{coef}.} The resulting tadpole graph corresponds to the 1-point conformal block. We assume that the loop is  more massive than the external leg so in this way we implement the perturbative procedure described in the previous section to obtain the linearized conformal block. To describe the deformation  we require that the radial deviation from the constant radius circle at $t=0$ is zero. Then, with time running from $0$ to $\pi$, the deviation reaches its maximum  value giving rise to the radial position of the vertex point where the external leg meets the loop.   

Recall that in the spherical case  the conical singularity/BTZ black hole is created by two heavy fields of the $n$-point  correlation function. In the toroidal case  the thermal AdS background is given a-priory, it is not created  by any fields of the 1-point  correlation function. Instead, the heavy object here is the  loop while the external leg is considered as producing small deformations. (The  loop can be loosely visualized as a potential line produced by two heavy fields at infinitely separated points of the cylinder  but with endpoints identified.) Nonetheless, both of them are treated as probe particles propagating on the thermal AdS background. \footnote{The configuration on Figure \bref{annulus} can be generalized by replacing the loop segment  by a geodesic winding around the non-contractible cycle. The resulting configuration is characterized by a winding number and should lead to other local minima of the gravitational action. The dual CFT interpretation is not obvious though.}

\subsection{Worldline approach} 
\label{sec:worldline}

The worldline approach proved to be useful in calculating lengths of various geodesic networks \cite{Hijano:2015rla,Alkalaev:2015wia,Banerjee:2016qca}. Each geodesic segment is described by the following action 
\be
\label{total2}
S = \int_1^2 d\lambda \sqrt{g_{mn}(x)\dot x^m \dot x^n}\;,
\ee
where $1$ and $2$ are initial/final positions, local coordinates are $x^m = (t,\phi,r)$, the metric coefficients $g_{mn}(x)$ are read off from \eqref{thermal}, and the velocity $\dot x$ is defined with respect to the evolution parameter  $\lambda$. The action is reparametrization invariant so that one can impose the normalization condition    
\be
\label{normal}
|g_{mn}(x)\dot x^m \dot x^n| = 1\;,
\ee
allowing one to identify the on-shell value of the action as $S = \lambda_2 - \lambda_1$ while the corresponding equations of motion describe geodesic curves in the thermal AdS space \eqref{thermal}. 

As the metric coefficients do not depend on time and angular variables it follows that the corresponding momenta are  conserved, $\dot p_t = 0$ and $\dot p_\varphi = 0$. We restrict the dynamics to the level surface characterized by the constant angle $\varphi =0$  corresponding to conserved $p_\phi = 0$ while the other conserved momentum $p_t$ is the motion integral giving  the shape of a geodesics curve. The level surface is identified with the annulus resulting from slicing the thermal AdS along the non-contractible cycle. The normalization condition is then $g_{tt}\, \dot t^2 + g_{rr}\, \dot r^2 = 1$, and taking into account  $g_{tt}g_{rr}  = -\tau^2$ we find out that 
\be
\label{dotr}
\dot r  = \pm \sqrt{r^2 -s^2+1}\;,\qquad\;\; s \equiv  i\, \frac{|p_t|}{\tau}\;.
\ee
The overall sign corresponds to the direction of the evolution parameter flow.

The circumference of the loop can be calculated by using the definition of the time momentum $p_t = g_{tt}\dot t$. Representing the loop as two semi-loops we find that   
\be
\label{sloop}
\dot t  =\frac{ i}{\tau}\,\frac{ s}{1+r^2}\;:\qquad S_{loop} = \frac{2\tau}{is}\int_0^{\pi} dt \big(1+r^2(t)\big)\;,
\ee
where $s$ is the loop momentum, and  the radial deviation $r(t)$ is defined from the normalization condition \eqref{dotr}. Note that the radial deviation is the distance between the blue loop and the dashed circle shown on Fig. \bref{annulus}. It satisfies the boundary condition
\be
\label{rho}
r(0) = 0\;, \qquad r(\pi) = \rho\;,
\ee
where $\rho$ is the vertex radial position (see our comments below). \footnote{One can also impose less restrictive condition $r(0) = \tilde\rho$ implying that the loop may have non-vanishing radial deviation in the upper point. We are grateful to Per Kraus for pointing us this possibility which will be considered elsewhere. } 
If the loop is a constant radius circle then from  \eqref{dotr} we find  that $s^2 = 1+r^2$, and, therefore, $S_{loop} = -2\pi i \tau  s$. For $r=0$ the length is $S_{loop} = -2\pi i  \tau$.

The time momentum of the external leg is shown below to be vanishing, $s = 0$. It means precisely that this line is stretched along the radial direction and has  the length given by
\be
\label{leg}
S_{leg} = \int_{\rho}^{\Lambda} \frac{dr}{\sqrt{1+r^2}} = -\text{ArcSinh}\,\rho\;,
\ee
where $\rho$ is the vertex radial position. The cutoff parameter $\Lambda$ is introduced to regularize the conformal boundary position. The integration term arising in the $\Lambda \to \infty$ limit is just an infinite constant to be discarded in the final result.

The boundary condition \eqref{rho} is pretty well justified in the low temperature regime $\im \tau \gg 1$ because in this limit the loop circumference  is large. The disturbance produced by the external leg in the vertex point  gradually decreases  when approaching the upper point of the loop $t=0$. This intuitive estimate will be confirmed  below by explicit calculations (see our comments below \eqref{rtexplicit}).

\subsection{Equilibrium condition}
\label{sec:equilibrium}

The configuration of the geodesic segments near the vertex point is described by the sum of three pieces of the type  \eqref{total2}. Minimizing such a combination we find that the time momenta at the vertex point \eqref{rho}  satisfy the weighted equilibrium condition 
\be
\label{econd}
\tilde\epsilon  p_m^1 + \tilde\epsilon  p_m^2 + \epsilon  p_m^0 = 0\;, 
\ee
where $p^{1,2}$ are the ingoing/outgoing intermediate momenta and $p^0$ is an external momentum at the vertex point. As the tadpole graph lies on the constant angle slice the spacetime index takes just two values $m = (t, r)$. Here, we suppose that the evolution parameter $\lambda$ is increasing  away from the vertex.   Obviously, any closed curve has $|p_m^1| = |p_m^2|$ while overall sings can be different. Indeed, their relative sign  $\pm$ depends on whether we take $m=t$ or $m=r$, see below.  

Using   parametrization \eqref{dotr} we find that the time component of \eqref{econd} is given by 
\be
\tilde \epsilon(s_1 - s_2) + \epsilon s_0 = 0\;.
\ee   
As the loop has $s_1 = s_2 \equiv s$ we find out that  $s_0  =0$. In other words, the boundary-to-bulk segment goes along the radial direction. The radial component of the equilibrium condition \eqref{econd} is more interesting
\be
\label{eqcon}
\tilde \epsilon (\dot r_1 + \dot r_2)  - \epsilon \dot r_0 = 0\;, 
\ee
where $\dot r$ are given by \eqref{dotr} with positive sign. Since $r_1 = r_2$ we find that 
$\delta \dot r_0  = 2 \dot r_1$, where $\delta = \epsilon/\tepsilon$. 
Then, the vertex position $\rho$ is expressed in terms of the loop momentum $s$ as
\be
\label{vertex}
\rho = \sqrt{\frac{s^2}{1-\delta^2/4}-1}\;\;.
\ee
In particular, we see that if the external field is decoupled $\delta = 0$ then there is the following solution
\be
\label{seed}
\delta = 0\;:\qquad s = 1\;, \qquad \rho = 0\;.
\ee 
This is the seed solution for the perturbation procedure that we develop in the next sections. It means that in this case the loop is a circle, while the vertex position $\rho =0$ and $t = \pi$ is fixed to be the point where the external field leg is going to be attached to.

\subsection{Half-cycle condition}
\label{sec:half}

To find how the loop radial deviation   evolves in time we use  the time loop momentum $p_t = g_{tt} \dot t$ and recall that $\dot r$ is given by \eqref{dotr}. Taking their ratio we arrive at the first order differential equation  
\be
\label{diff}
\frac{dt}{dr} = \frac{i}{\tau} \frac{s}{(1+r^2)\sqrt{r^2 - s^2+1}}\;,
\ee 
where $s$ is the loop momentum. In the range  $t\in [0,t]$ and $r\in [0,r]$ the above equation integrates to  
\be
\label{rt}
e^{-2i \tau t}(r^2+1)(s^2-1)   = -r(2 s \sqrt{r^2-s^2+1}+r s^2+r)+s^2-1\;,
\ee
where the time dependence is given only via the exponential. Note that the modular parameter enters only through this exponential and this is how the elliptic parameter \eqref{qtau} appears in the final expressions. Solution \eqref{rt} can be used to find the vertex position \eqref{rho} through the half-cycle  condition represented as 
\be
\label{half}
e^{-2i\pi \tau }   = \frac{-\rho(2 s \sqrt{\rho^2-s^2+1}+\rho s^2+\rho)+s^2-1}{(\rho^2+1)(s^2-1)}\;,
\ee
which is given by substituting $t = \pi$ and $\rho = r(\pi)$ into \eqref{rt}. We note that values \eqref{seed} solve the above equation.

The vertex condition \eqref{vertex} and the half-cycle condition \eqref{half} are two algebraic equations on two variables  $\rho$ and $s$. The system can be solved unambiguously as $ s = s(\delta, \tau)$ and $\rho = \rho(\delta, \tau)$. However, contrary to the vertex condition we see that the half-cycle condition is quite complicated which makes it problematic to find  exact solutions. Instead, in the next section we develop the deformation method which allows one to find $s$ and $\rho$ as a power series in $\delta$ with the leading term given by the seed solution \eqref{seed}.

\subsection{Perturbative expansion} 
\label{sec:perturbative}

Following the deformation method in \cite{Alkalaev:2015wia,Alkalaev:2015lca} we consider the tadpole graph perturbatively starting from the seed solution corresponding to the loop of constant radius \eqref{seed} and adding small interaction with the external leg. In this way we are led to calculate the length function 
\be
\label{Stot}
L =  \tepsilon\, S_{loop}(\tau, \epsilon, \tepsilon)  + \epsilon\, S_{leg}(\tau, \epsilon, \tepsilon)\;,
\ee
where $S_{loop}$ and $S_{leg}$ are given by \eqref{sloop} and \eqref{leg}. It is obvious that the loop and radial segments are parameterized by the modular and conformal  parameters. In particular, both lengths can be  represented as power series in small parameter $\delta$ introduced in \eqref{lightness} as
\be
\label{powerser}
\ba{c}
\dps
S_{loop} = \sum_{n=0}^\infty  S^{(n)}_{loop}(\tau) \delta^n\;,
\\

\dps
S_{leg} = \sum_{n=0}^\infty  S^{(n)}_{leg}(\tau) \delta^n\;,

\ea
\ee
where $S^{(0)}_{loop}(\tau) = -2\pi i \tau$  and $S^{(0)}_{leg}(\tau) =0$. Noting that $\tilde \epsilon \delta^n = \epsilon \delta^{n-1}$  the  length function is given by 
\be
\label{Sdec}
L =  -2\pi i \tilde \epsilon \tau  + \tepsilon\,\sum_{n=1}^\infty  \Big[ S^{(n)}_{loop}(\tau)+S^{(n-1)}_{leg}(\tau)\Big] \delta^n\;.
\ee
Comparing with the linearized block expression \eqref{vovalin} we find out that  the following condition is to be satisfied 
\be
\label{odd}
S^{(n)}_{loop}(\tau)+S^{(n-1)}_{leg}(\tau)=0\;, \qquad  \text{for}\qquad  n=2k+1\;, \quad k = 0,1,2,...\;.
\ee
Below we show that this property of the expansion coefficients follows from the particular form of the time momentum and the vertex position as functions of $\delta$.

Now, both the time momentum and the radial deviation along with the vertex position are analogously expanded as 
\be
\label{rspert}
s= \sum_{n=0}^\infty s_n \delta^n \;,
\qquad
r(t)= \sum_{n=0}^\infty r_n(t) \delta^n\;,
\qquad
\rho= \sum_{n=0}^\infty \rho_n \delta^n\;,
\ee
where the seed values are $s_0=1$, $r_0(t) = 0$ and $\rho_0 = 0$, cf. \eqref{rho}. 
It follows that the loop length \eqref{sloop} is 
\be
\label{loopP}
\ba{l}
\dps
S_{loop}= -2\pi i\tau + 2\pi i \tau   s_1 \delta -2 i \tau \delta ^2 \int_0^{\pi}dt \big( r_1^2(t)+ s_1^2- s_2\big)
\\
\\
\dps
\hspace{45mm}-2 i \tau \delta^3 \int_0^{\pi}dt\big(-r_1^2(t) s_1+2 r_1(t) r_2(t)- s_1^3+2  s_1 s_2- s_3\big)+ \cO(\delta^4)\;.
\ea
\ee
Analogously, the radial length \eqref{leg} is expanded as 
\be
\label{radP}
S_{leg} = -  \rho_1\delta - \rho_2\delta^2- (\rho_3-\frac{\rho_1^3}{6})\delta^3+\cO(\delta^4)\;. 
\ee

\vspace{-5mm}

\paragraph{Perturbative momentum and radius.} In what follows we solve the evolution equation \eqref{rt} and the vertex equation \eqref{vertex} perturbatively. It turns out that being expanded in $\delta$ the evolution equation contains half-integer order terms  $\delta^{m/2}$ for $m=1,2,3,... \,$. Solving the evolution equation in this case gives rise to $0/0$ terms. However, all  indeterminate forms can be avoided provided that the expansion \eqref{rspert} has $s_{2k+1} = 0$ and $r_{2k} = 0$  for any $k=0,1,2,... \,$. It follows that the modified expansions read
\be
\label{rssolv}
\ba{l}
\dps
s= \sum_{n=0}^\infty s_{2n} \delta^{2n}  = s_0+ s_2 \delta^2 + s_4 \delta^4+...\;,
\\
\dps
r(t)= \sum_{n=0}^\infty r_{2n+1}(t) \delta^{2n+1} =  r_1(t) \delta+r_3(t)\delta^3+... \;,

\\
\dps
\rho= \sum_{n=0}^\infty \rho_{2n+1} \delta^{2n+1}  = \rho_1 \delta + \rho_3 \delta^3+...\;.

\ea
\ee
In particular, inspecting \eqref{loopP} and \eqref{radP} we find that the condition \eqref{odd} is now automatically satisfied. 

The lowest order terms of the vertex  equation and the evolution equations are given by 
\be
 (1-2\sqrt{\rho_1^2-2 s_2})\delta+\cO(\delta^2) = 0\;, 
\ee
and 
\be
-2 \delta^2 \big((\sqrt{r_1^2-2s_2}+r_1)r_1 +s_2 (e^{-2i \tau   t}-1)\big)+\cO(\delta^3) = 0\;.
\ee
Solving them in each order in $\delta$ we find first radial corrections  
\be
\label{rtexplicit}
\ba{l}
\dps
r_1(t) = \frac{1}{2} \text{sech}\left(-i\pi  \tau\right) \sinh \left(-i\tau  t\right)\;,
\\
\\
\dps
r_3(t) = \frac{1}{16} \text{sech}^3\left(-i\pi  \tau \right) \sinh \left(-i\tau  t\right) \cosh ^2\left(-i\tau  t\right)\;,
\\
\\
\dps
r_5(t) = \frac{3}{256} \text{sech}^5\left(-i\pi  \tau \right) \sinh \left(-i\tau  t\right) \cosh ^4\left(-i\tau  t\right)\;, ...\;,
\ea
\ee 
and the loop momentum corrections
\be
\label{sexplicit}
\ba{l}
\dps
s_2 = -\frac{1}{8} \text{sech}^2\left(-i\pi \tau \right)\;,
\quad
s_4=-\frac{1}{128} \text{sech}^4\left(-i\pi  \tau\right)\;,
\quad
s_6=-\frac{1}{{1024}}\text{sech}^6\left(-i\pi  \tau\right)\;, ....\;.
\ea
\ee
The dependence on $\tau$ shows that for $\im \tau \approx 1$ the radial deviation quickly starts to increase near the upper point of the loop, while for large $\im \tau \gg 1$ it is supported  near the vertex point.   

Substituting $t = \pi$ into \eqref{rtexplicit} we find the vertex position corrections  
\be
\label{verexplicit}
\rho_1 = \frac{1}{2} \tanh \left(-i\pi  \tau \right)\;,
\quad
\rho_3 = \frac{1}{16} \tanh \left(-i\pi  \tau\right)\;,
\quad
\rho_5 = \frac{3}{256} \tanh \left(-i\pi  \tau \right)\;, ...\;.
\ee
The appearance of the common factor in the above formulae  allows one to represent the vertex position as
\be
\label{rhoexact}
\rho = g(\delta) \tanh \left(-i\pi  \tau\right)\;, \qquad \text{where}\qquad
g(\delta) = \half \delta +\frac{1}{16} \delta^2+ \frac{3}{256}\delta^3+...\;.
\ee  
It follows that the characteristic dependence on $\tau$ is retained while higher orders in $\delta$  appear as the overall scale factor. Also, since the modular parameter is pure imaginary  the radial position is not periodic in time $t \sim t + 2\pi$. Finally, let us note that there is another branch of radial corrections with overall minus sign. It corresponds to the tadpole graph in the inner region on Fig. \bref{annulus}.     

\vspace{-3mm}

\paragraph{Perturbative lengths.} Using corrections \eqref{rtexplicit}, \eqref{sexplicit}, and \eqref{verexplicit} it is straightforward to find corrections to the loop length and the leg length according to \eqref{loopP} and  \eqref{radP}. Consider first the lowest order nontrivial corrections
\be
S^{(2)}_{loop}(\tau) = -\frac{1}{4} \tanh \left(-i\pi  \tau \right) \;,\qquad S^{(1)}_{leg}(\tau) = \frac{1}{2} \tanh \left(-i\pi  \tau \right)\;.
\ee  
Identifying the modular parameters as 
\be
\label{modular}
\hat \tau  = \tau+\half \;, 
\ee 
we find  that the above functions in the conformal parameterization  are exactly  
\be
S^{(2)}_{loop}(q) :=  f_1(q)+\frac{1}{4}\;,\qquad S^{(1)}_{leg}(q) := -2 f_1(q) - \half\;,
\ee
and so  modulo an additive constant the  length function  \eqref{Stot}, \eqref{Sdec} in  the first nontrivial order  is  
\be
L(q):=  -\tepsilon \log q  -\tepsilon\,\delta^2 f_1(q)+ \cO(\delta^4)\;.
\ee
Adding the thermal AdS action term and comparing then with the first coefficients of the linearized classical block \eqref{vovalin} we find out that $S_{thermal}(q)+L(\epsilon, \tepsilon|q) = -f(\epsilon, \tepsilon|q)$, cf. \eqref{identification}. Higher order corrections can be taken into account quite analogously. Assuming \eqref{modular} one can show that up to multiplicative and additive constants  $S^{(2k)}_{loop}  \sim f_k$ and $S^{(2k-1)}_{leg}  \sim f_k$ so that their sought-for combination is $S^{(2k)}_{loop}+ S^{(2k-1)}_{leg} = -f_k + const_k$, and the identification \eqref{identification} holds true. 

Finally, recall that in the low-temperature regime,  where the thermal AdS dominates the functional integral, the imaginary part is $Im(\tau)\gg 1 $ so that the real part $1/2$ in \eqref{modular} can apparently be ignored. It follows that in this limit the modular parameters coincide and therefore we deal with the same torus. 

\section{Conclusion}
\label{sec:conclusion}

In this paper we have proposed the AdS/CFT interpretation of the 1-point linearized classical conformal block on a torus. We have found that the block function can be calculated as  the geodesic length of  the tadpole graph embedded into the thermal AdS space plus the holomorphic part of the thermal AdS action. 

We have seen that the correspondence holds in the low-temperature regime $\im \tau \gg 1$, where the gravity functional integral is dominated by the thermal AdS. Then, the loop circumference is large  so that  its upper point can be treated  as a point at infinity with respect to the vertex point where the leg is attached to the loop. This justifies our choice of the boundary condition  \eqref{rho} which says that the radial deviation of the deformed loop from the zeroth radius circle vanishes at the upper point. Moreover, examining the radial deviation as a function of $\tau$ we find out that the disturbance created by the external leg is indeed localized near the vertex point  when $\im \tau \gg 1$. Otherwise, the disturbance creates non-vanishing radial deviation already near the upper point. 

On the other hand, it is known that  there are other saddle points dominating the gravity functional integral at other temperatures or other $\tau$ \cite{Maldacena:1998bw}. In particular, in the high-temperature regime $\im \tau \ll 1$ the thermal AdS has less action than BTZ black hole with the modular parameter related to that of the thermal AdS by the modular transformation $\tau \to -1/\tau$. However, the worldline interpretation of the linearized classical  1-point toroidal block function $f^{lin}(\epsilon, \tepsilon|\hat \tau)$ is specific to the thermal AdS and not BTZ black hole just because the leading term here is $\sim \hat \tau$ rather than $\sim 1/\hat\tau$.  


In particular, it explains why the toroidal CFT and the thermal AdS modular parameters cannot be equal to each other for any temperatures. Instead,  they are related as in \eqref{modular}, that is the thermal AdS modulus $\tau$ is shifted by $1/2$ with respect to the CFT  parameter $\hat\tau$ while  in the low-temperature approximation this difference is negligibly small. It is also worth noting that the geodesics length is a real function of real parameters so that the conformal block function should  also be real. It implies that the elliptic parameter satisfies the reality condition $q = \bar q$ so that up to the modular equivalence there are two branches with $\text{Re}\, \hat\tau = 0,\,1/2$ and arbitrary $\im \hat\tau$. Excluding the first branch we are naturally left with the second one realized via the holographic calculations.

Let us  conclude  by discussing some interesting future perspectives. The natural extension of the present results would be  to develop holographic interpretation of $2$-point and higher-point conformal toroidal blocks, including the $W_N$ symmetry case \cite{deBoer:2014sna}.  Also, it is important to study the subleading $1/c$ corrections and related phenomena along the lines of \cite{Beccaria:2015shq,Fitzpatrick:2015dlt,Anous:2016kss}. 
The holographic consideration of various entropies in CFT on a torus is also interesting \cite{Herzog:2013py,Datta:2013hba,Datta:2014zpa}.  It would be especially important to understand the bulk worldline interpretation of the toroidal conformal blocks in terms of Wilson lines within the Chern-Simons formulation \cite{deBoer:2013vca,Ammon:2013hba}, for recent discussion see, \textit{e.g.}, \cite{Hegde:2015dqh,Bhatta:2016hpz,Melnikov:2016eun,kraus_new}. Finally, it is interesting to develop the geodesic Witten diagrams technique by analogy with conformal blocks on the complex plane \cite{Hijano:2015qja}. In particular, a dual interpretation of the OPE coefficients in the toroidal case, and, more generally, how the operator algebra is realized in terms of the bulk theory is yet to be found.

\vspace{7mm} 

\noindent \textbf{Acknowledgements.} We thank V. Didenko and, especially, P. Kraus for  useful discussions. The work of K.A. was supported by the Russian Science Foundation grant 14-42-00047 in association with Lebedev Physical Institute. The work of V.B. was performed with the financial support of the Russian Science
Foundation (Grant No.14-50-00150).

\appendix

\section{The linearized block coefficients }
\label{sec:appA}

The $q$-expansion of the coefficients \eqref{vovalin} is given by 
\begin{align}
\label{appcoef1}
&f_1(q)=\frac{1}{2} q  (1+q+q^2+q^3+q^4+q^5+q^6+q^7+q^8+q^9+q^{10}+...)\;,\\
&f_2(q)=-\frac{1}{48} q^2 (3+8 q+15 q^2+24 q^3+35 q^4+48 q^5+63 q^6+80 q^7+99 q^8+120 q^9+...)\;,\\
&f_3(q)=\frac{1}{480} q^3 (10+45 q+126 q^2+280 q^3+540 q^4+945 q^5+1540 q^6+2376 q^7+...)\;,\\
&f_4(q)=-\frac{1}{3584} q^4 (35+224 q+840 q^2+2400 q^3+5775 q^4+12320 q^5+24024 q^6+...)\;,\\ 
&f_5(q)=\frac{1}{23040}q^5 (126+1050 q+4950 q^2+17325 q^3+50050 q^4+126126 q^5+...)\label{appcoef2}\;.
\end{align}

\providecommand{\href}[2]{#2}\begingroup\raggedright
\addtolength{\baselineskip}{-3pt} \addtolength{\parskip}{-1pt}


\providecommand{\href}[2]{#2}\begingroup\raggedright\endgroup

\end{document}